\documentclass[final,5p,times,twocolumn]{elsarticle}

\usepackage{amssymb}

\journal{arXiv}

\usepackage{xspace}
\usepackage{csquotes}
\usepackage[labelfont=bf]{caption}
\usepackage{enumitem}
\usepackage{newtxtext,newtxmath}
\usepackage{siunitx}
    \DeclareSIUnit{\point}{point}
\usepackage[dvipsnames]{xcolor}
\usepackage[most]{tcolorbox}
\usepackage{multicol}
\usepackage{tikz}
    \usetikzlibrary{trees}
\usepackage{appendix}

\usepackage[acronym]{glossaries}

\usepackage{hyperref}
\hypersetup{pdfauthor=author,colorlinks=true}  

\newacronym{eam}{EAM}{electro-anatomical mapping}
\newacronym{ecg}{ECG}{electro-cardiogram}
\newacronym{egm}{EGM}{electrogram}
\newacronym{csp}{CSP}{conduction system pacing}
\newacronym{cdt}{CDT}{cardiac digital twin}
\newacronym{cep}{CEP}{cardiac electrophysiology}
\newacronym{vt}{VT}{ventricular tachycardia}
\newacronym{lat}{LAT}{local activation time}
\newacronym{lge}{LGE}{late gadolinium enhanced}
\newacronym{scn}{SCN}{Spatial-Configuration Net}

\newcommand{\oC}{\emph{\href{https://opencarp.org}{\mbox{openCARP}}}\xspace}
\newcommand{\paraview}{\emph{\href{https://www.paraview.org}{\mbox{ParaView}}}\xspace}
\newcommand{\cU}{\emph{\href{https://git.opencarp.org/openCARP/carputils}{\mbox{carputils}}}\xspace}
\newcommand{\pE}{\emph{pyCEPS}\xspace}
\newcommand{\studio}{\emph{CARPentry Studio}\xspace}
\newcommand{\cartoMap}{\emph{1-LV Sinus}\xspace}

\newcommand{\clinic}{University Hospital Graz, Graz, Austria\xspace}
\newcommand{\carto}{Carto3\xspace}
\newcommand{\navx}{enSite Precision\xspace}
\newcommand{\oEP}{\emph{openEP}\xspace}
\newcommand{\python}{Python~3\xspace}

\begin{document}

\begin{frontmatter}



\title{\pE: A cross-platform Electroanatomic Mapping Data to Computational Model Conversion Platform \\ 
for the Calibration of Digital Twin Models of Cardiac Electrophysiology}


\author[inst1]{Robert Arnold}
\author[inst1]{Anton J. Prassl}
\author[inst2]{Aurel Neic}
\author[inst1,inst3]{Franz Thaler}
\author[inst1]{Christoph M. Augustin}
\author[inst1]{Matthias A.~F. Gsell}
\author[inst1,inst4]{Karli Gillette}
\author[inst5]{Martin Manninger}
\author[inst5]{Daniel Scherr}
\author[inst1,inst2,inst4]{Gernot Plank\corref{cor1}}
\ead{gernot.plank@medunigraz.at}

\cortext[cor1]{Corresponding author}

\affiliation[inst1]{
    organization={Gottfried Schatz Research Center, Division of Medical Physics and Biophysics, Medical University of Graz},
    city={Graz},
    country={Austria}}
\affiliation[inst2]{
    organization={NumeriCor GmbH},
    city={Graz},
    country={Austria}}
\affiliation[inst3]{
    organization={Institute of Computer Graphics and Vision, Graz University of Technology},
    city={Graz},
    country={Austria}}
\affiliation[inst4]{
    organization={BioTechMed-Graz},
    city={Graz},
    country={Austria}
    }
\affiliation[inst5]{
    organization={Division of Cardiology, Department of Medicine, Medical University of Graz},
    city={Graz},
    country={Austria}}

\begin{abstract}
\textbf{Background and Objective:} Data from \gls{eam} systems are playing an increasingly important role in computational modeling studies 
for the patient-specific calibration of digital twin models. 
However, data exported from commercial \gls{eam} systems are challenging to access and parse.
Converting to data formats 
that are easily amenable to be viewed and analyzed with commonly used cardiac simulation software tools such as \oC remains challenging. 
We therefore developed an open-source platform, \pE, for parsing and converting clinical \gls{eam} data conveniently to standard formats widely adopted within the cardiac modeling community.\\
\textbf{Methods and Results:} \pE is an open-source Python-based platform providing the following functions: 
(i) access and interrogate the \gls{eam} data exported from clinical mapping systems; 
(ii) efficient browsing of \gls{eam} data to preview mapping procedures,  \glspl{egm}, and \glspl{ecg}; 
(iii) conversion to modeling formats according to the \oC standard, 
to be amenable to analysis with standard tools and advanced workflows as used for \emph{in silico} \gls{eam} data. 
Documentation and training material to facilitate access to this complementary research tool for new users is provided.
We describe the technological underpinnings and demonstrate the capabilities of \pE first, 
and showcase its use in an exemplary modeling application where we use clinical imaging data to build a patient-specific anatomical model.\\
\textbf{Conclusion:} 
With \pE we offer an open-source framework for accessing \gls{eam} data,
and converting these to cardiac modeling standard formats.
\pE provides the core functionality needed to integrate \gls{eam} data in cardiac modeling research.
We detail how \pE could be integrated into model calibration workflows facilitating the calibration of a computational model based on \gls{eam} data.
\end{abstract}



\begin{keyword}
electro-anatomical mapping \sep computational modeling \sep cardiac digital twins \sep model calibration


\end{keyword}

\end{frontmatter}


\section{Introduction}
\label{sec:introduction1}
Computational modeling and simulation of \gls{cep}
has been playing a pivotal role in basic cardiology research~\cite{Niederer-2018-ID12269} since decades
as a unique approach for gaining insight into mechanisms to explain experimental and clinical observations.
Significant advances in modeling and imaging technologies made over the past decade
have rendered anatomically accurate and biophysically detailed \emph{in silico} models 
of human \gls{cep} at the organ scale feasible.
These models show high promise as a clinical research tool, for device and drug development~\cite{fda2016}, and even as a complementary clinical modality. 

When used as a clinical modality, it could aide in diagnosis, therapy stratification, and planning in future precision cardiology~\cite{Corral-Acero-2020-ID13459}.
Of particular relevance for clinical therapy planning are \gls{cep} models
that are calibrated to uniquely replicate like-for-like all available clinical observations for a given patient. 
Such high fidelity models -- referred to as \glspl{cdt}~\cite{Corral-Acero-2020-ID13459} -- may offer predictive potential to be utilized to predetermine the prognosis of acute therapeutic responses. 

A major challenge in using \glspl{cdt} is the calibration of the model 
to observable clinical data of a given patient.
Beyond the clinical standard 12-lead \gls{ecg}, \gls{eam} data is the most routine and accurate data available of a patient's \gls{cep}.
\gls{eam} data are always acquired during ablation procedures and, increasingly often, during other procedures such as device implants
in therapies for example in \gls{csp}~\cite{wang_2023:_eam_csp}.

While \gls{eam} data are abundantly available clinically, 
they are not easily accessible to be used for the calibration of \glspl{cdt}  
for numerous reasons. In terms of data visualization, quick browsing and data previewing of \gls{eam} data requires access to clinical software that is, in general, not conveniently available 
within the wider cardiac modeling community. Immediate processing of \gls{eam} data with computational modeling tools is limited as data exported from \gls{eam} systems are stored in proprietary formats for which reading and parsing can be challenging and time-consuming to implement. Inherent problems with spatio-temporal registration 
between computational model and \gls{eam} data also pose an issue. 

In this work we address these issues by introducing the Python-based software \pE~\cite{arnold_2024_10606459} providing the following functions: 
(i) access and interrogate the \gls{eam} data exported from \gls{eam} systems;
(ii) efficient browsing of \gls{eam} data to preview mapping procedures, body surface \gls{ecg} recordings, and intracardially recorded \glspl{egm}; 
(iii) conversion to modeling formats according to the \oC standard, 
to be amenable to analysis with the same tools and processing workflows 
as used for \emph{in silico} \gls{eam} data.
One exemplary data set acquired with the \carto \gls{eam} system is provided for testing, 
along with documentation and training material to facilitate access to this complementary research tool for new users. 
We describe the technological underpinnings and demonstrate the capabilities of \pE first, 
and showcase its integration in a simple exemplary modeling application where clinical imaging data is used to build a patient-specific anatomical model
and inform the \gls{cep} of the model using \pE converted \gls{eam} data.

\section{Materials and Methods}
\label{sec:methods}

\subsection{Data Structure and Implementation}  
The \pE software in its current implementation provides capabilities to access \gls{eam} data from \carto \gls{eam} systems,
preview the entire range of available data types 
and export and convert these to standard formats used in the cardiac modeling community using the \oC simulator and the \cU framework.
The available data types include anatomical manifolds, measurement point clouds, \gls{egm}-derived parameter maps, and \gls{ecg} and \gls{egm} data.
The data is stored by \gls{eam} systems in proprietary data formats,
which are read in by \pE first,
and are then saved to disk as single binary serialized object file (\enquote{pickling}).
A hierarchical data structure is used within \pE (see \ref{app:datastructure}, Fig.~\ref{fig:data_structure}) that
mirrors the hierarchical structure of \gls{eam} data sets within \carto.
Namely, each exported data set from an \gls{eam} system represents a study, i.e.\ an interventional procedure.
During each procedure, multiple mapping sequences are performed for example during sinus rhythm, pacing or an arrhythmia, or after ablation.
Each map is then constructed from multiple \gls{egm} recordings at varying locations distributed over the region of interest.
Locations at which radio-frequency ablation was performed and where the energy was delivered are recorded and stored.
This information is combined with metrics quantifying the ablation performance.

The implementation of \pE was developed based on \python to ascertain usability on a broad range of compute platforms
and to ease adoption among the target user groups, i.e.\ the computational modeling community.
Routinely used functions to import, export and view data are accessible via a command line interface.
The software basis of \pE has been implemented from scratch, 
with a clearly defined application focus,
to keep the code basis compact and, thus, to facilitate continuous development with limited resources. 
The \pE software described here is made available under the \href{http://www.gnu.org/licenses/gpl.html}{GNU General Public License 3 (GPLv3)}. 

\subsection{Clinical Test Data}
For developing and testing of \pE, mapping data sets exported from a \carto \gls{eam} system were used.
Typically, such data sets consist of a large number of files (10k+) 
that contain all information recorded during the mapping procedure.
An XML file describing the overall study encodes the organization of data
that is required for processing the data set (included maps, map-associated data such as locations of recording sites, filenames of associated data, ...).
All files associated with a study are exported by \carto as ZIP compressed archive.
\pE can import studies from extracted ZIP archives but also enables import directly from ZIP archives.

The \carto study chosen to showcase \pE capabilities in this work 
stems from a patient suffering from recurrent \gls{vt} episodes who was treated by \gls{vt} ablation at the \clinic. 
The exported ZIP archive had a size of \qty{\sim 2}{\giga\byte} (unzipped archive \qty{\sim 22}{\giga\byte}) 
and consisted of \num{\sim 36000} files. 
Three mapping procedures were performed: 
one during sinus rhythm (\num{2282} mapping points), 
one during \gls{vt} (\num{1728} mapping points), and, 
one after ablation (\num{63} mapping points). 
To terminate \gls{vt} \num{43} ablation sites were used.
\gls{eam} data were acquired from patients included in the local ablation registry. This registry was approved by the ethics committee of the Medical University of Graz (reference number 31-036 ex 18/19).

\subsection{Performance Benchmarking}
To benchmark the performance of \pE, two metrics were considered: 
(i) the time required to read clinical \gls{eam} data and convert it to the \pE data structure, and, 
(ii) the file size of the generated serialized data representation.
Both metrics were compared to the performance measured with \oEP~\cite{williams2021:_openep} implemented in MATLAB. 
Data from \num{21} \carto studies were exported in compressed ZIP format and read by \pE directly from the ZIP archive.
All mapping procedures with all recording points and ablation sites were automatically imported.
The generated \pE data structure was saved to a file, and file size was determined using \python standard modules.
Additionally the time required to convert the \pE data structure in its entirety to \oC compatible file formats, 
as described below in Sec.~\ref{sec:_conversion}, was measured.
This metric also included the time taken to export 12-lead \gls{ecg} data 
which is not part of the automatically generated \pE data structure and, therefore, 
had to be additionally read from the clinical \gls{eam} data set.
Performance benchmarks were carried out on a standard desktop PC (HP ProDesk, Intel i7-6700 @ \qty{3.4}{\giga\hertz}, \qty{32}{\giga\byte} RAM, \qty{2}{\tera\byte} HDD storage) using Spyder IDE 5.4.3 and Python\ 3.8.10.

\subsection{Data Browsing and Visualization}
\label{sec:visualization}
After importing \gls{eam} studies and creating \pE data objects 
these can be conveniently visualized to quickly assess data quality and gain an overview of the procedure.
Visualization was implemented in \verb|dash| and \verb|dash_vtk| which creates a local HTML page opened in the browser, 
thus, facilitating the platform-independent visualization of studies.
The viewer offers browsing of all included mapping data:
The acquired anatomical manifold can be visualized along with all recording points and/or ablation sites. Body surface \gls{ecg} traces and intracardially recorded \gls{egm} traces for individual recording points are shown.
\Gls{egm}-derived surface parameter maps interpolated onto the anatomical manifold such as \gls{lat} and 
uni- or bipolar voltage maps are either shown as exported along with \gls{eam} data or parameters are interpolated onto the anatomical manifold using an inverse-distance-weighting algorithm to provide a spatial view.
Body surface \glspl{ecg} (Einthoven, Goldberger, Wilson) are recorded during the entire mapping procedure. 
Since mapping points are acquired sequentially the body surface \glspl{ecg} associated to individual recording points vary, 
depending on the respective acquisition time the mapping point was recorded.
To obtain body surface \glspl{ecg} that are representative for an entire mapping procedure three methods were implemented.
Method \textbf{M1} calculates the median amplitude at each point of time, 
method \textbf{M2} selects a recorded signal with the lowest mean-squared-error within a window-of-interest when compared to the median representation obtained by \textbf{M1}, and
method \textbf{M3} selects a recorded signal with the highest cross-correlation within a window-of-interest compared to the median representation obtained by \textbf{M1}.

\subsection{Data Conversion to \oC Modeling Formats}
\label{sec:_conversion}
While clinically acquired mapping data (location of recording points, triangulated meshes representing anatomical manifolds, multi-channel time signals, etc.) are structurally similar to simulated data (nodes, meshes, spatio-temporal signals) they are stored in proprietary file formats that are optimized for analysis and display with commercial \gls{eam} systems, but are incompatible with commonly used cardiac simulation or visualization software such as \oC or \paraview. 
Therefore, to use clinical \gls{eam} data in modeling applications two key issues have to be addressed: 
(i) data must be converted to data formats
that are compatible with tools used in simulation workflows;
and,
(ii) spatio-temporal registration of \gls{eam} data with computational anatomies and simulated data is required.
\pE addresses the first issue by offering convenient support for format conversion. 
Either, individual subsets of data can be selected to be read and converted, or, all contained data are imported, and exploded into a set of files that are readable using \oC tools.
Converted data are stored and presented by \pE in a systematic and coherent manner, ensuring easy comprehension and accessibility (see Fig.~\ref{fig:output_folder}).

For each mapping procedure, \pE reads and converts the following types of data, using labels in the filename to indicate the data type:
\begin{itemize}
    \renewcommand\labelitemi{--}
    \setlength\itemsep{0em}
    \item \textbf{\textit{.surf.}} : anatomical manifolds
    \item \textbf{\textit{.bsecg.}} : body surface \glspl{ecg} representative for the mapping procedure
    \item \textbf{\textit{.ecg.}} : time traces of the 12-lead \glspl{ecg} for each recording point
    \item \textbf{\textit{.egm.}} : time traces of the unipolar, bipolar and reference \glspl{egm} for each recording point
    \item \textbf{\textit{.map.}} : \gls{egm}-derived surface parameter maps interpolated onto the anatomical manifold
    \item \textbf{\textit{.ptdata.}} : \gls{egm}-derived parameters for each recording point
    \item \textbf{\textit{.lesions.}} : Ablation procedure data
\end{itemize}
Exported data have different forms and dimensions and are converted to files in the following formats, with suitable file extensions:
\begin{itemize}
    \renewcommand\labelitemi{--}
    \setlength\itemsep{0em}
    \item \textbf{\textit{.pts}} : Cartesian coordinate triples with $P$ entries for nodal location of anatomical surface meshes, $R$ entries for recording point positions, or with $L$ entries for ablation catheter positions
    \item \textbf{\textit{.elem}} : Triangular elements of anatomical surface meshes with $E$ triples of values
    \item \textbf{\textit{.dat}} : ASCII-encoded data vectors of dimension $R$ for data corresponding to recording points, 
    dimension $P$ for interpolated data onto the anatomical manifold, and dimension $L$ for data corresponding to ablation sites
    \item \textbf{\textit{.igb}} : Binary-encoded spatio-temporal data of dimension $T \times R$ where $T$ is the number of sampling points acquired over time
\end{itemize}
Alternatively, export to a \verb|*.vtk| container format, where both meshes and data are combined, is also supported.
Detailed information on input and output formats are available in the \href{https://opencarp.org/documentation/user-manual}{\oC~\cite{openCARP-paper} user manual} (see Secs.~4.1 and 4.2). 
An exception are ASCII time traces for the body surface  \glspl{ecg}, stored as \verb|*.json| files. 
This format is well supported in Python but is not yet referenced as \oC format.

Data corresponding to recording points can be viewed on a point cloud representing the actual recording sites (\verb|*.pc.pts|), or on a projected point cloud where each recording location is projected onto the closest position at the anatomical manifold (\verb|*.ppc.pts|).
Such data is indicated in the \pE export with the suffix \verb|*.pc.*|.
Note, that \pE, by default, only exports valid data points, 
i.e.\ data points for which the \gls{eam} system deemed the recording site to be in sufficiently close vicinity to the anatomical manifold, and where the recording position was stable over the recording time window.

\subsection{Computational Model Building}
The use of \pE is demonstrated in an exemplary modeling workflow.
A patient suffering from recurrent \gls{vt} episodes was treated by \gls{vt} ablation at the \clinic.
Prior to the ablation procedure, contrast CT and \gls{lge} MRI data were acquired, and during \gls{vt} ablation therapy comprehensive \gls{eam} data were recorded. 
A previously developed image-to-anatomical model generation workflow was used to build an anatomically accurate model of the patient's heart and torso~\cite{crozier15}. 
Image segmentation of the contrast CT was obtained using the deep learning-based \gls{scn}~\cite{payer2019integrating,payer2017multi}, which was trained on an in-house dataset.
Before being processed by the SCN, the original resolution of the contrast CT (0.39 mm in-plane, 0.7 mm out-of-plane) was resampled to an isotropic resolution of 1.5 mm due to memory constraints during training.
The obtained segmentation was resampled to the original resolution and verified by an expert.
A volumetric biventricular mesh~\cite{prassl09:_tarantula} of the segmented CT images along with rule-based ventricular fiber arrangements~\cite{bayer12} and universal ventricular coordinates~\cite{BAYER201883} was generated then using \studio (NumeriCor GmbH, Graz, Austria) at an average resolution of \SI{1}{\milli \meter} which is suitable for fast simulations using a reaction-eikonal electrophysiology model~\cite{neic2017:_reaction_eikonal}.

\subsection{\glsentrytext{eam} Data Integration with Computational Model}
\Gls{eam} data recorded during an intervention and the computational anatomical model built from imaging data are defined in two different reference coordinate systems.
Integration of \Gls{eam} data with a computational model requires therefore 
(i) a spatial transformation 
such that measured and simulated data live in the same space, and a spatial correlation between measured and simulated data points can be established, and,
(ii) a transfer operator must be defined for projecting measured data onto a region of interest of the model, and \emph{vice versa}. 
Spatial registration can be rigid, keeping the spatial relation between recording locations fixed, or non-rigid, where spatial relation may be altered to some extent, to obtain a better overlap between measurements at a point cloud with the computational model.
For the sake of this study, we opted for a rigid affine transformation, $\mathbf{M}\in\mathbb{R}^{4\times4}$, to register the anatomical manifold with the left ventricular endocardium.
The same transformation $\mathbf{M}$ is applied then to all point data in the data set.
Shepard's interpolation method~\cite{Shepard1968} was used to transfer data measured on the recording point cloud onto the anatomical manifold, and onto the matching endocardial manifold of the computational model.
Local minima of the \gls{lat} map were identified and used as earliest activation sites to compute an activation sequence using a reaction-eikonal model of ventricular electrophysiology~\cite{neic2017:_reaction_eikonal}.
A lead field approach was used to compute \glspl{egm} at selected recording sites.

\section{Results}
\label{sec:results}
\pE provides a command line interface for most commonly used functions. In the following sections the commands used are given in verbatim. For detailed usage and examples see the project homepage on \href{https://pypi.org/project/pyCEPS/}{PyPI} or the software repository on \href{https://github.com/medunigraz/pyCEPS}{GitHub}.

\subsection{Data Parsing}
\label{sec:parsing}
\begin{figure}[h!tb]
    \centering
    {\includegraphics[width=0.75\columnwidth]{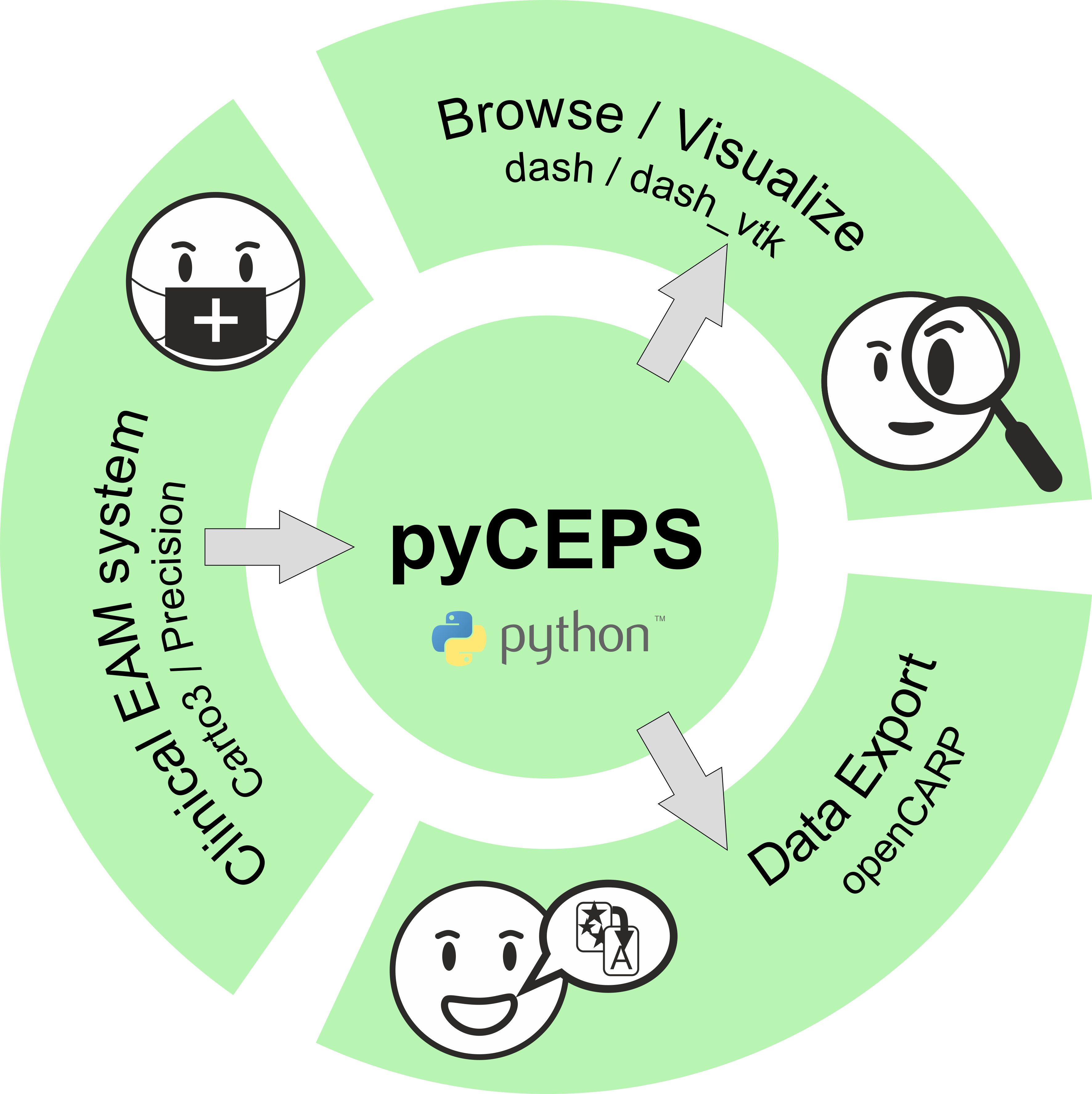}}
    \caption{\textbf{\pE overview:} \pE is used to parse and translate data exported from clinical \gls{eam} systems, quickly browse cases, and export data to \oC compatible file formats, or other common formats used in the cardiac modeling community.}
    \label{fig:overview}
\end{figure}

An overview of the \pE architecture is shown in Fig.~\ref{fig:overview}.
\pE can import studies from extracted ZIP archives but also enables import directly from ZIP archives to avoid the cumbersome and disk space consuming task of extracting data first.
Importing \gls{eam} data starts by specifying the clinical system (\verb|--system|) and the location of the \gls{eam} data set  (\verb|--study-repository|).
For \carto studies \pE automatically locates the XML file describing the study and lists all available mapping procedures within the study.
A particular or all available mapping procedures can subsequently be imported (\verb|--import-map|).
Importing of all relevant data is automated and a \pE data structure is created which can be saved (\verb|--save-study|) as binary serialized object file.
Previously generated serialized objects files can be loaded by \pE (\verb|--pkl-file|) and additional data can be imported and added at later time.

\subsection{Performance Benchmark}
\begin{figure*}[h!tb]
    \centering
    {\includegraphics[width=0.75\textwidth]{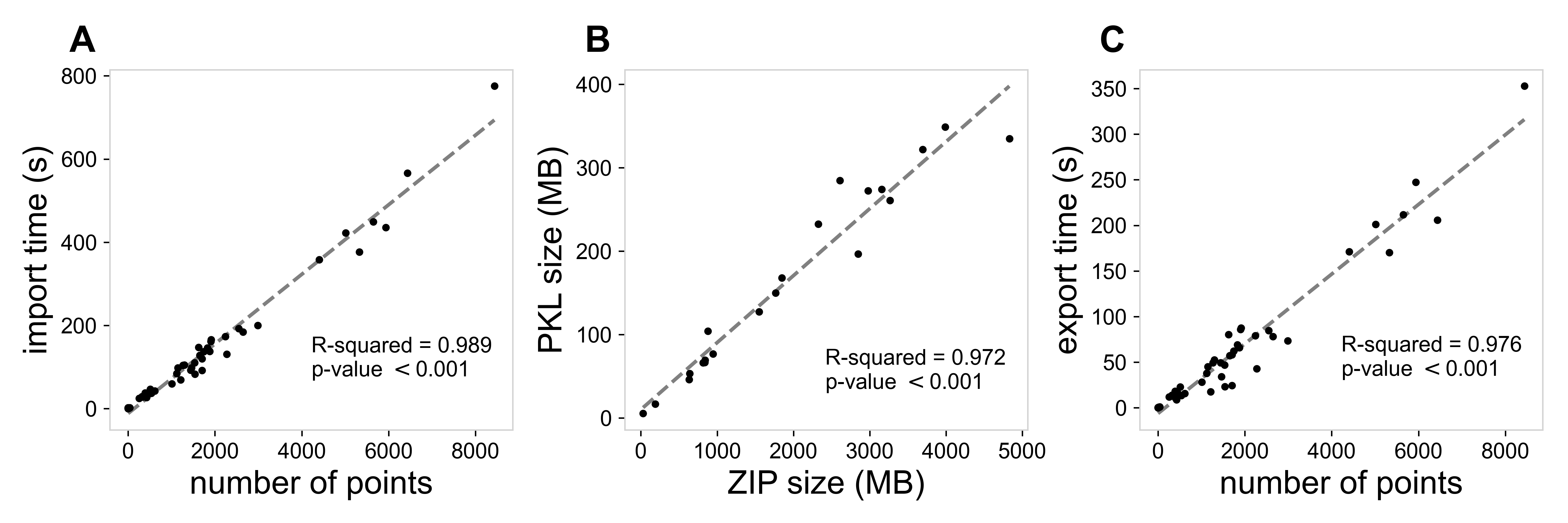}}
    \caption{\textbf{Performance benchmark: } \textbf{(A)} Import time is increasing with number of recording points (\qty[per-mode = symbol]{\approx 84}{\milli\second\per\point}).
    \textbf{(B)} Storage size of serialized data objects as a function of the original exported \gls{eam} data. 
    Disk space required was by a factor \num{\approx 12} smaller than the exported raw data.
    \textbf{(C)} Exporting serialized data objects to \oC compatible formats is increasing with the number of recording points (\qty[per-mode = symbol]{\approx 38}{\milli\second\per\point}). These export timings include the loading of additional \gls{ecg} data per recording point.}
    \label{fig:benchmark_import_time}
\end{figure*}

The data set for performance benchmarking comprised \num{21} \carto studies with \num[separate-uncertainty = true]{3771(2475)} recording points per mapping procedure (range \numrange[range-phrase=--]{62}{8854}).
The exported ZIP archives from the \gls{eam} system had a size of \qty[separate-uncertainty = true, separate-uncertainty-units=single]{1.73(1.05)}{\giga\byte}.
Data import took \qty[separate-uncertainty = true, separate-uncertainty-units=single]{101(63.7)}{\second}. Import time was increasing with number of recording points with a linear regression slope of \qty[per-mode = symbol]{84}{\milli\second\per\point} (Fig.~\ref{fig:benchmark_import_time}A).
This is approximately by a factor \num{5} faster than other available tools like the MATLAB-based \oEP~\cite{williams2021:_openep}.
The generated serialized data objects had a size of \qty[separate-uncertainty = true, separate-uncertainty-units=single]{149.8(96.4)}{\mega\byte} which was significantly smaller than the exported data from the \gls{eam} system.
Data size was reduced approximately by a factor of \num{12} (Fig.~\ref{fig:benchmark_import_time}B).
Exporting data to \oC compatible formats took \qty[separate-uncertainty = true, separate-uncertainty-units=single]{45.8(32.2)}{\second}. Export timings  also include loading additional \gls{ecg} data per recording point which is not included during standard import of data sets (Fig.~\ref{fig:benchmark_import_time}C).
All values are given as median $\pm$ median absolute deviation (MAD).

\subsection{Data Browsing and Visualization}
\begin{figure*}[h!tb]
    \centering
    {\includegraphics[width=0.95\textwidth]{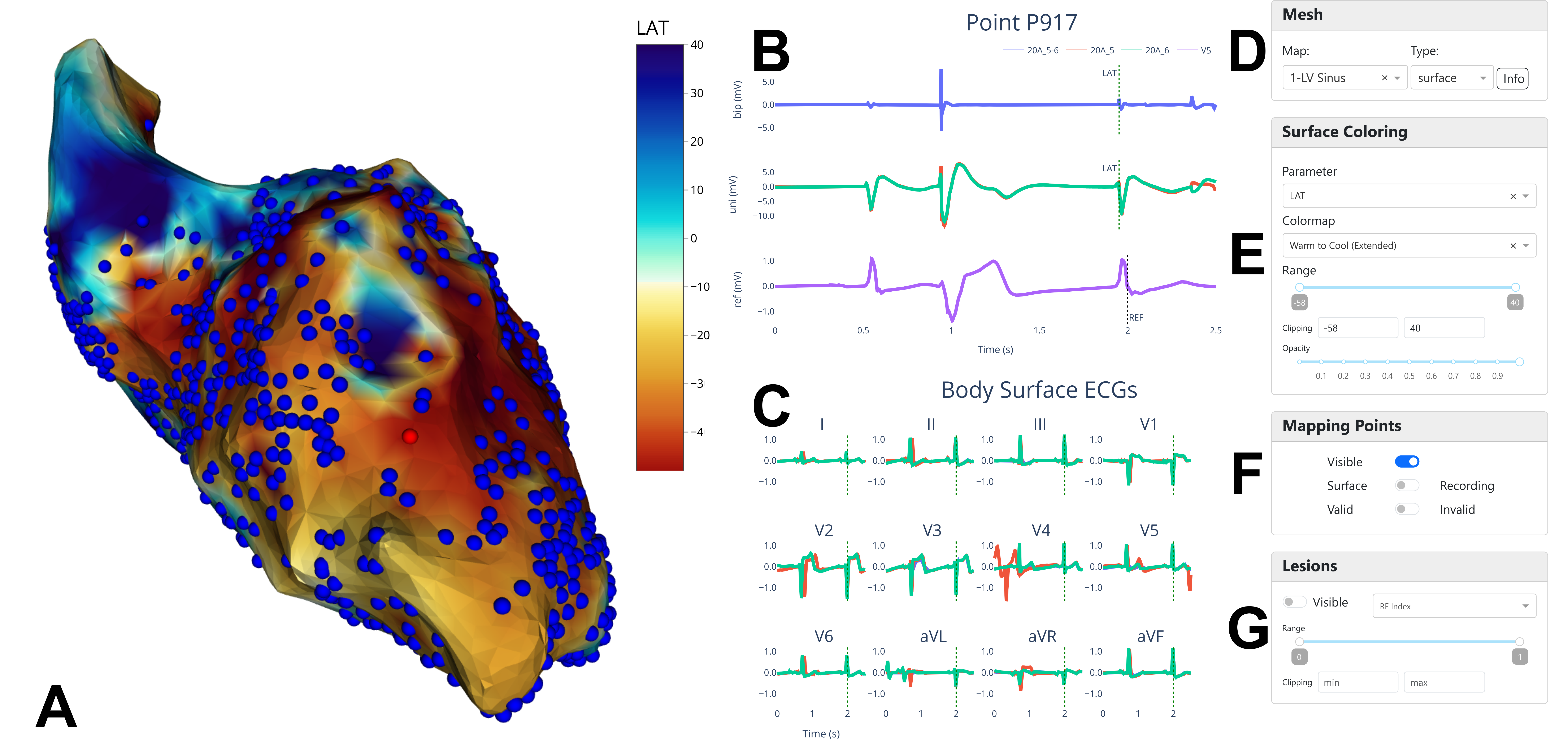}}
    \caption{\textbf{Data browsing capabilities of \pE:} \textbf{(A)} Shown are the anatomical manifold model acquired before the procedure, the electrode positions at which \glspl{egm} were recorded (blue dots), and a color coded parameter map interpolated from sparse discrete recording sites upon the anatomical manifold. \textbf{(B)} For a user selected recording site unipolar and bipolar \glspl{egm} are shown. \textbf{(C)} Representative 12-lead body surface \glspl{ecg} for the mapping procedure are shown. \textbf{(D)} Selection of mapping procedure. \textbf{(E)} Selection and adjustment of surface parameter maps. \textbf{(F)} Toggling and adjustment of recording points features. \textbf{(G)} Toggling and adjustment of ablation site features.
    }
    \label{fig:dash_interface}
\end{figure*}

After selecting a \pE serialized data object for visualization with
\begin{verbatim}
    pyceps --system 'carto' \
           --pkl-file 'PATH/TO/STUDY.PKL' \
           --visualize
\end{verbatim}
the user can browse all mapping procedures included in the data object.
Capabilities for quick browsing of \gls{eam} data are illustrated in Fig.~\ref{fig:dash_interface}.
The user can visualize the electro-anatomical manifold approximating the chamber geometry along with recorded \gls{egm} and \gls{ecg} data, interpolated surface parameter maps, and ablation site data.
All included mapping procedures in the \pE data object are listed and can be selected (Fig.~\ref{fig:dash_interface}D).
The chamber geometry for the selected mapping procedure is shown (Fig.~\ref{fig:dash_interface}A) and voltage and activation maps to be shown on the chamber geometry can be selected and customized (Fig.~\ref{fig:dash_interface}E).
Locations where \gls{egm} data was recorded can be shown 
either at their actual recording position 
or projected onto the closest point on the high-resolution chamber geometry (Fig.~\ref{fig:dash_interface}F).
Recording points (blue dots in Fig.~\ref{fig:dash_interface}A) can be selected by the user to visualize recorded \gls{egm} time traces for the selected recording site (Fig.~\ref{fig:dash_interface}B).
Ablation lesions can be added and are shown with a color-coded ablation index (Fig.~\ref{fig:dash_interface}G).

\begin{figure*}[h!tb]
    \centering
    {\includegraphics[width=0.75\textwidth]{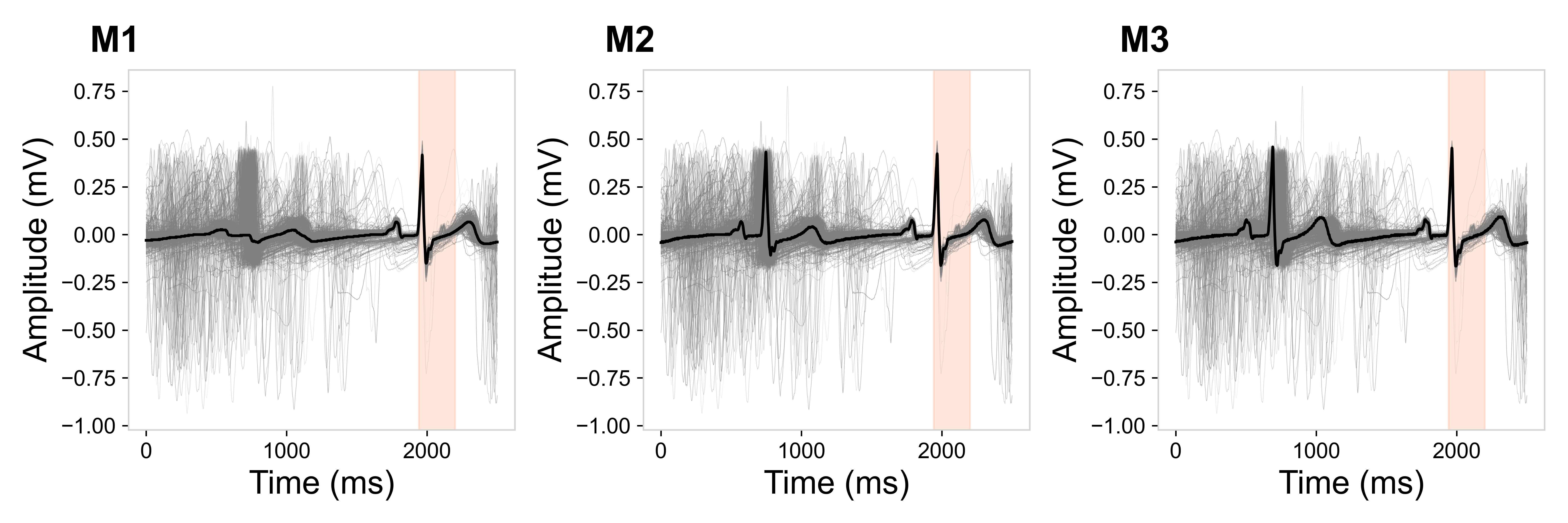}}
    \caption{\textbf{Representative body surface ECGs:}
    Three methods to obtain representative body surface \glspl{ecg} on the example of Einthoven\ I are shown.
    Gray traces represent the bundle of \glspl{ecg} that were recorded for each measurement point during sequential \gls{egm} acquisition.
    Black traces are the representative \glspl{ecg} for the given mapping procedure.
    \textbf{(M1)} Median value at each point of time.
    \textbf{(M2)} Recorded signal with lowest mean-square-error.
    \textbf{(M3)} Recorded signal with highest cross-correlation.
    Note that method M1 yields an artificial signal while methods M2 and M3 yield a signal that was actually recorded (see trace in M1 over the time window \qtyrange[range-phrase = --]{0}{1500}{\milli\second}).
    The window-of-interest used for methods M2 and M3 is highlighted for all methods.}
    \label{fig:bsecg}
\end{figure*}

For each mapping procedure consisting of sequentially recorded points,  representative body surface \gls{ecg} traces are shown (Fig.~\ref{fig:dash_interface}C) that were determined in three different ways as described in Sec.~\ref{sec:visualization}.
Note that the \enquote{median method} \textbf{M1} yields an artificial signal 
while both other methods yield an actually recorded (real) signal 
that best matches the artificial median signal.
The resulting representative \gls{ecg} traces are illustrated in Fig.~\ref{fig:bsecg} for the Einthoven I lead.

\subsection{Data Conversion to \oC Modeling Format}
\begin{figure*}[h!tb]
    \centering
    {\includegraphics[width=.75\textwidth]{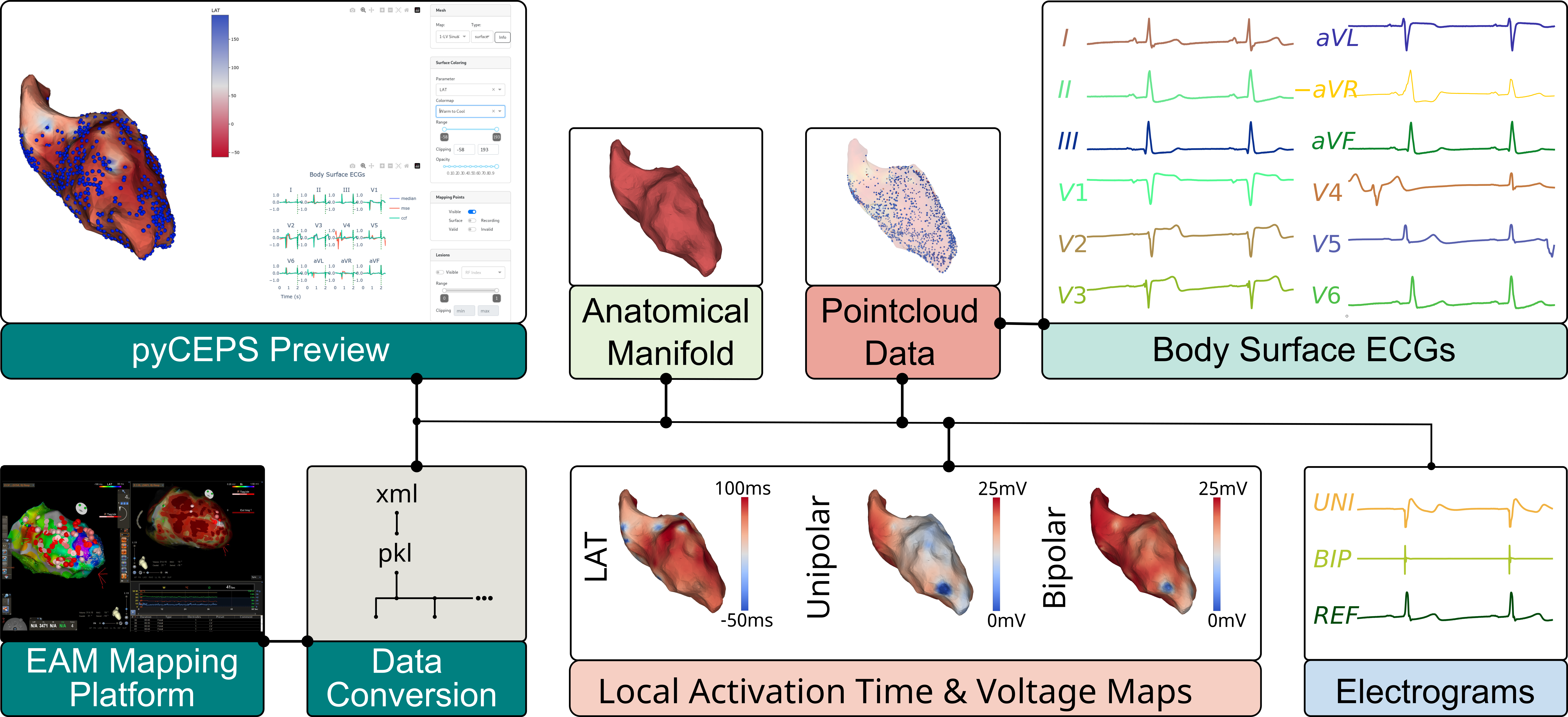}}
    \caption{\textbf{Data Conversion:} Workflow for exporting \gls{eam} data using \pE.
    \Gls{eam} data are exported as ZIP archives and converted to a serialized data structure ({\tt .pkl} file).
    \pE is used for previewing data in the converted study, and for exporting data to standard formats used in the cardiac modeling community.
    These include \oC formatted surface meshes representing the anatomical mapping manifold, recorded and projected point clouds with recorded \glspl{egm} and derived parameters (\gls{lat}, uni- and bipolar voltage magnitudes), interpolations of the point cloud data onto the anatomical manifold, and body surface  \glspl{ecg}.
    }
    \label{fig:dataconversion}
\end{figure*}

After importing \gls{eam} data and saving a serialized data structure (see Sec.\ \ref{sec:parsing}), the data structure can be converted to \oC compatible data formats.
The name of a mapping procedure is given (\verb|--map|) followed by a list of data items to convert by specifying the parameters:
\begin{itemize}
    \setlength\itemsep{0em}
    \item \verb|--dump-point-egms|
    \item \verb|--dump-point-ecgs|
    \item \verb|--dump-point-data|
    \item \verb|--dump-surface-maps|
    \item \verb|--dump-map-ecgs|
    \item \verb|--dump-lesions|
    \item \verb|--dump-mesh|.
\end{itemize}
These parameters obtain \glspl{egm}, \glspl{ecg} (recorded during point acquisition), \gls{egm}-derived parameters, interpolated surface maps, body surface \glspl{ecg} (representative for the mapping procedure), ablation sites, and anatomical geometries, respectively, when called individually.
Example usage would therefore be
\begin{verbatim}
    pyceps --system 'carto'
           --pkl-file 'PATH/TO/STUDY.PKL' \
           --map 'MAP_NAME' \
           --dump-mesh
\end{verbatim}
For convenience, the function \verb|--convert| is available, which imports a complete \gls{eam} data set and exports all of the above mentioned data items automatically.
The standard workflow to import, visualize, and convert data from a \carto \gls{eam} system is shown in Fig.~\ref{fig:dataconversion} and would be executed on the console with
\begin{verbatim}
    pyceps --system 'carto' \
           --pkl-file 'PATH/TO/STUDY.PKL' \
           --convert \
           --visualize \
           --save-study \
\end{verbatim}
All converted data pertaining to a mapping procedure are stored in sub-directories named after the mapping procedure.
Each converted dataset includes then: 
(i) a triangulated anatomical shell acquired initially during the procedure, 
(ii) unipolar and bipolar voltage maps and local activation time maps, 
(iii) unipolar, bipolar and reference electrograms, 
(iv) a 12-lead electrocardiogram, and, 
(v) location of ablation sites.
The directory tree of a converted mapping procedure is showcased in Fig.~\ref{fig:output_folder}.

\subsection{Integrating \glsentrytext{eam} data with a computational modeling workflow}

\begin{figure*}[h!tb]
    \centering
    {\includegraphics[width=0.75\textwidth]{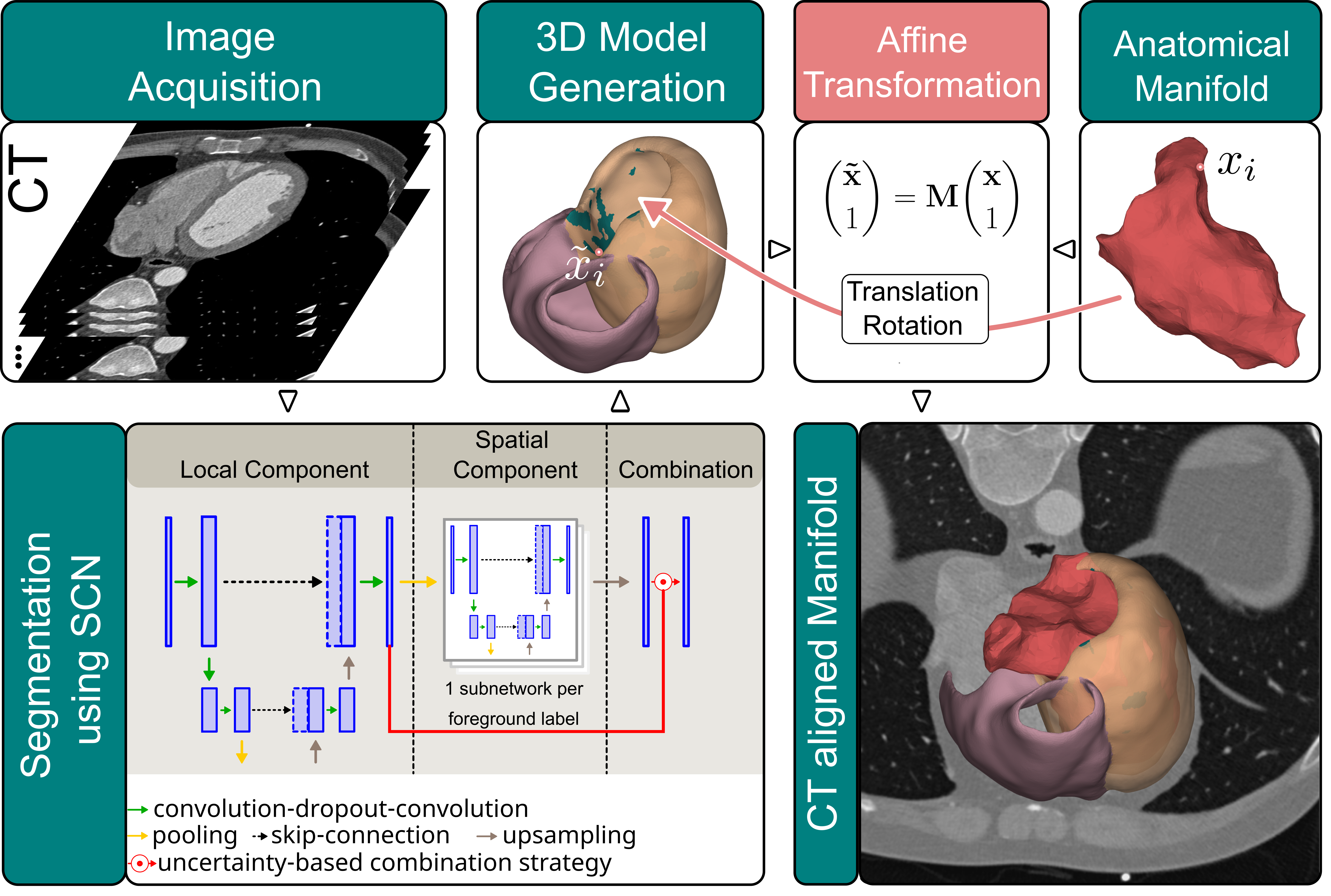}}
    \caption{\textbf{Model generation and alignment:} Image segmentation of the contrast CT was obtained using a deep learning-based \gls{scn}.
    A 3D finite element model of the ventricles was generated within \studio.
    An affine transformation to align the anatomical manifold to the left ventricular cavity of the 3D mesh was identified manually and stored for registering available \carto output into the spatial coordinates system of the image stack.}
    \label{fig:modelgen_and_alignment}
\end{figure*}

\begin{figure*}[h!tb]
    \centering
    {\includegraphics[width=.95\textwidth]{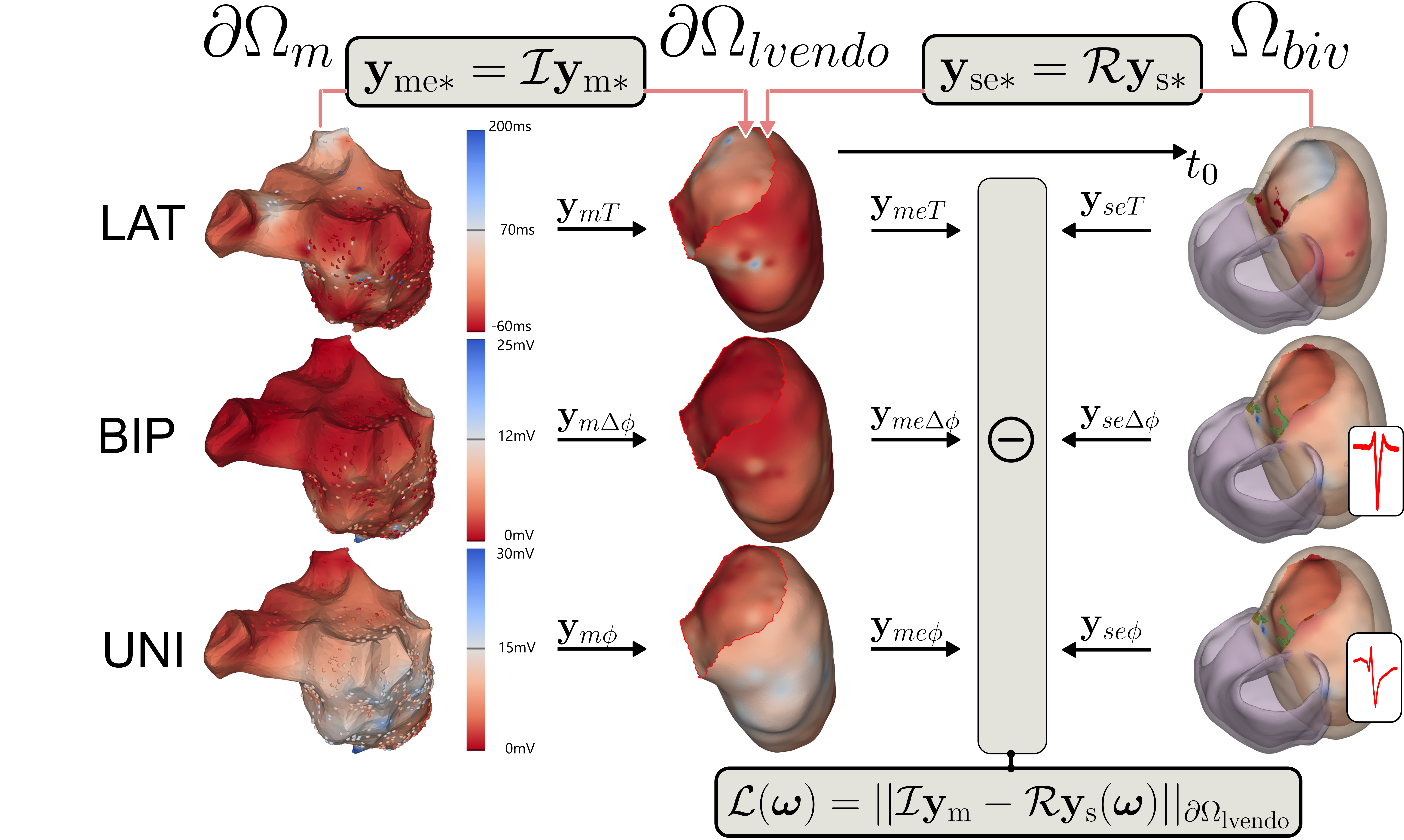}}
    \caption{\textbf{Mapping and interpolation:}
    The maps of \gls{lat}, $T$, unipolar and bipolar voltage magnitudes, $\phi$ and $\Delta \phi$, are interpolated from the \gls{eam} manifold $\partial\Omega_{\mathrm{m}}$ onto mesh points of the corresponding endocardial surface $\partial\Omega_{\mathrm{lvendo}}$ using an interpolation operator, $\mathcal{I}$, based on Shephard's method. Maps defined on $\partial\Omega_{\mathrm{lvendo}}$ are related to the 3D volumetric mesh $\Omega_{\mathrm{biv}}$ by the restriction operator $\mathcal{R}$. Thus, surface $\partial\Omega_{\mathrm{lvendo}}$ serves as a common reference for comparing data simulated over $\Omega_{\mathrm{biv}}$, where the difference between measured and simulated data can be expressed as a loss function $\mathcal{L}(\omega)$.}
    \label{fig:mapping_and_interpolation}
\end{figure*}

The integration of \pE in a representative modeling workflow (see Figs.~\ref{fig:modelgen_and_alignment} and \ref{fig:mapping_and_interpolation}) is showcased in an exemplary study of building a computational model of an ischemic cardiomyopathy patient treated by \gls{vt} ablation therapy.
The tomographic image stack was used as a spatial reference for generating the model. 
After automated segmentation of both ventricles~\cite{payer2017multi}, a computational biventricular  mesh was generated at a target resolution of \SI{1.2}{\milli \metre}~\cite{prassl09:_tarantula, crozier15},
and equipped with rule-based fibers~\cite{bayer12} and an anatomical reference frame~\cite{BAYER201883}.
All recorded \gls{eam} data comprising two mapping procedures (under normal sinus rhythm and during ongoing \gls{vt}) were converted to \oC modeling formats using \pE.
The \gls{eam} anatomical shell in model format was registered to the left ventricular endocardial surface of the 3D volumetric model,
defined as a subset of the full 3D volumetric biventricular mesh, 
in the image space using an affine transform, $\mathbf{M}\in\mathbb{R}^{4\times4}$, 
which was defined interactively using \studio (Numericor GmbH, Graz, Austria) (see Fig.~\ref{fig:modelgen_and_alignment}).
The resulting transformation $\mathbf{M}$ was stored
and applied to all point clouds in the \gls{eam} data set 
to preserve spatial relation with the recorded anatomical shell.

The \gls{egm}-derived surface parameter maps of \gls{lat}, $T(\mathbf{x}_{\rm r})$, unipolar and bipolar voltage magnitudes, $\phi(\mathbf{x}_{\rm r})$ and $\Delta \phi(\mathbf{x}_{\rm r})$,
represented by the nodal data vectors, $\mathbf{y}_{\rm mT}$, $\mathbf{y}_{\rm m\phi}$ and $\mathbf{y}_{\rm m\Delta \phi}$,
were then interpolated onto mesh points of the corresponding endocardial surface, $\partial\Omega_{\rm{lvendo}}$ 
to obtain the maps $T(\mathbf{x}_{\rm s})$, $\phi(\mathbf{x}_{\rm s})$, $\Delta \phi(\mathbf{x}_{\rm s})$, with $\mathbf{x}_{\rm r}$ defined on the recorded anatomical manifold and $\mathbf{x}_{\rm s}$ defined on the computational mesh respectively.
The interpolation operator, $\mathcal{I} : \partial \Omega_{m} \rightarrow \partial \Omega_{\mathrm{lvendo}}$, was built based on Shepard's interpolation method~\cite{Shepard1968}. 
Thus, the measured data vector of each map, $\mathbf{y}_{\rm m}$, is represented on the endocardial surface 
of the reference computational mesh by $\mathbf{y}_{\rm me} = \mathcal{I}\mathbf{y}_{\rm m}$.
These maps defined on $\partial\Omega_{\rm{lvendo}}$ are related to the 3D volumetric mesh $\Omega_{\rm biv}$
by the restriction operator, $\mathcal{R} : \Omega_{\rm{biv}} \rightarrow \partial\Omega_{\rm{lvendo}}$, which restricts simulated data defined on the entire computational domain, $\Omega_{\rm{biv}}$,
to the endocardial manifold, $\partial \Omega_{\rm{lvendo}}$.
Thus, simulated data on the reference domain  $\partial \Omega_{\rm{lvendo}}$ are found by $\mathbf{y}_{\rm se} = \mathcal{R} \mathbf{y}_{\rm s}$, 
or measured data are on $\partial\Omega_{\rm lvendo}$ inserted into $\Omega_{\rm biv}$ by the injection 
$\mathbf{y}_{\rm s} = \mathcal{R}^\top \mathbf{y}_{\rm se}$.
The surface $\partial \Omega_{\rm lvendo}$ serves now as a common reference for comparing data simulated over $\Omega_{\rm biv}$
with data measured on $\partial \Omega_{\rm m}$ where the difference between measured and simulated data can be expressed now as
\begin{equation*}
    \mathcal{L}(\boldsymbol{\omega}) = || \mathcal{I} \mathbf{y}_{\rm m} - \mathcal{R} \mathbf{y}_{\rm s}(\boldsymbol{\omega})||_{\partial \Omega_{\rm lvendo}}.
\end{equation*}
With all inter-grid transfer operators in place, the \gls{lat} map $\mathbf{y}_{\rm mT}=T(\mathbf{x}_{\rm r})$ 
is transferred to the computational mesh $\Omega_{\rm biv}$ by $(R^\top \circ \mathcal{I}) \mathbf{y}_{\rm m}$
with $(R^\top \circ \mathcal{I}) : \partial \Omega_{\rm m} \rightarrow \Omega_{\rm biv}$, 
and local minima of the \gls{lat} map are used as boundary values for a reaction-eikonal model~\cite{neic2017:_reaction_eikonal}
to drive a biventricular activation sequence.
\Glspl{egm} in the blood pool of the left ventricle were computed using a lead-field approach~\cite{gillette21:_ep_twin}.
The differences in endocardial activation sequences and in \glspl{egm} are illustrated in Fig.~\ref{fig:mapping_and_interpolation}.
This constitutes only a first step of model calibration problem where model parameters $\boldsymbol{\omega}$ must be inferred 
by minimizing the loss $\mathcal{L}(\boldsymbol{\omega})$.

\section{Discussion}
\label{sec:discussion}
The use of clinical \gls{eam} data in computational modeling studies relies on the ability to read and browse the proprietary datasets produced by clinical \gls{eam} systems, 
and to register these datasets with a computational model, 
to facilitate a comparison between clinical observations and computed model predictions.
In this study we report on a new software for these tasks. 
Our implementation of \pE supports reading and parsing \gls{eam} data sets 
as generated by two widely used clinical \gls{eam} systems,
\carto (Biosense Webster) and \navx (Abbott).
Basic visualization capabilities using a standard web browser are built in to allow for a quick review and inspection
of clinical data sets.
The main functionality of \pE is the conversion from proprietary \gls{eam} data formats 
to standard modeling data formats as used within the \oC community, 
to facilitate a seamless integration within standard processing workflows used in the computational modeling domains.
\pE is implemented in \python, the most widely used language in the cardiac modeling community for data analysis, and 
is available in source code under an open-source license 
\href{http://www.gnu.org/licenses/gpl.html}{(GPLv3)},
along with basic documentation and one example data set.
Current functionality of \pE is restricted on data format conversion.
However, being implemented in \python, \pE is readily extensible 
to be used for data analysis in both modeling and clinical research using \gls{eam} data. 
We demonstrate the use of \pE in one simple exemplary application of the software where
we showcase the integration of \pE in a clinical modeling workflow 
for modeling an ischemic cardiomyopathy patient undergoing \gls{vt} ablation.

\subsection{Related software}
Reading and viewing clinical \gls{eam} data sets relies mostly on commercial software 
provided by the vendors of clinical \gls{eam} systems, which is not readily available outside of clinical cardiology departments.
Clinical research on \gls{eam} data cannot be carried out within these closed \gls{eam} software ecosystem, 
but is supported by options to export data in various formats. 
In clinical research carried out by cardiology fellows simpler formats readily accessible 
to be read and analyzed with MATLAB are preferred. 
Research software supporting the systematic processing of entire \gls{eam} datasets is scarce.
A notable exception is the software \oEP~\cite{williams2021:_openep} which fills this gap.
\oEP targets clinical research on \gls{eam} data analysis, 
and is implemented in MATLAB as the most common data analysis software used in the clinical research domain.
Specifically, \oEP also offers capabilities for reading and parsing of clinical \gls{eam} data sets, 
it implements advanced data analysis algorithms, and supports a range of visualization techniques. 
However, in contrast to \pE, \oEP is clearly more geared towards clinical users
with interest in advanced data analysis of clinical data,
and not for computational modeling communities mainly interested in making clinical data sets accessible 
in modeling applications.
The main focus of \pE is the conversion of data into formats 
that are accessible with the many advanced data processing tools developed within the modeling communities
for analyzing simulated data. Specifically, \pE converts \gls{eam} data seamlessly to \oC formats
that are compatible with the various workflows implemented in the \oC modeling ecosystem.
Further, \pE is written in \python which is much more widely used and adopted in the cardiac modeling community 
whereas \oEP is implemented in MATLAB which might be more convenient to use in clinical research but is not free of charge.
In principle, both approaches \pE and \oEP can be used interchangeably, with \oEP offering additional \gls{eam} analysis methods
such as more advanced data interpolation and analysis techniques, 
while \pE is more tailored for being integrated in a modeling workflow.
However, as many data analysis algorithms for \gls{eam} data sets have been developed in Python~\cite{lubrecht2021:_piemap,costabal2020:_pinn}
these are straight forward to integrate in \pE to expand the data analysis capabilities.

\subsection{Limitations}
In its current implementation \pE best supports the handling of \gls{eam} data acquired with \carto systems,
as the vast majority of data sets used for testing were acquired with this system.
Our access to Abbott \navx data was rather limited. As such \pE for \navx was tested only with two small data sets.
Thus, not all features could be fully implemented and extensively tested.
Advanced data analysis tools for \gls{eam} data as used in clinical research are missing.

\section{Conclusion}
\label{sec:conclusion}
With \pE we offer an open-source framework for parsing and viewing of \gls{eam} data,
and converting these to cardiac modeling standard formats as used within the \oC community.
Thus, \pE provides the core functionality needed to integrate \gls{eam} data in cardiac modeling research, 
by comparing clinical observations to model predictions.
Advanced \gls{eam} data analysis tools such as the regularization of \gls{lat} maps~\cite{lubrecht2021:_piemap},
the computation of conduction velocity maps~\cite{padilla_2023:_conduction_velocity_mapping}, 
rotor~\cite{jones_2013:_rotor_mapping} 
or dominant frequency analysis~\cite{sanders_2005:_df_mapping},
or the inference of fiber architecture~\cite{costabal2020:_pinn} are missing.
However, \pE is implemented in \python which is most widely used in the cardiac modeling community, 
and therefore readily facilitates the extension of \pE with more advanced \gls{eam} data analysis tools.

\section*{CRediT authorship contribution statement}
\textbf{Robert Arnold:} Conceptualization, Methodology, Software, Validation, Formal analysis, Writing - Original Draft, Writing - Review \& Editing, Visualization.
\textbf{Anton J. Prassl:} Conceptualization, Methodology, Software, Validation, Formal analysis, Writing - Original Draft, Writing - Review \& Editing, Visualization.
\textbf{Aurel Neic:} Formal analysis, Writing - Review \& Editing.
\textbf{Franz Thaler:} Methodology, Formal analysis, Writing - Review \& Editing.
\textbf{Christoph M. Augustin} Writing - Review \& Editing.
\textbf{Matthias A.~F. Gsell:} Methodology, Formal analysis, Writing - Original Draft, Writing - Review \& Editing.
\textbf{Karli K. Gillette:} Methodology, Formal analysis, Writing - Original Draft, Writing - Review \& Editing.
\textbf{Martin Manninger-Wünscher:} Investigation, Resources, Data Curation, Writing - Review \& Editing.
\textbf{Daniel Scherr:} Investigation, Resources, Data Curation, Writing - Review \& Editing.
\textbf{Gernot Plank:} Conceptualization, Methodology, Validation, Formal analysis, Resources, Writing - Original Draft, Writing - Review \& Editing, Supervision, Project administration, Funding acquisition.

\section*{Acknowledgments}
This research was funded in part by the Austrian Science Fund (FWF) [I-6476-B] and the Austrian Research Promotion Agency (FFG) [FO999891133].

\section*{Declaration of competing interest}
The authors declare that they have no known competing financial/commercial interests, or personal relationships that could have appeared to influence the work reported in this paper.

 \bibliographystyle{elsarticle-num} 
 \bibliography{pyceps}






\onecolumn
\appendix

\counterwithin{figure}{section}
\counterwithin{table}{section}
\renewcommand\thefigure{\thesection\arabic{figure}}
\renewcommand\thetable{\thesection\arabic{table}}

\section{\pE Data Structure}
\label{app:datastructure}

\tikzstyle{every node}=[draw=black,thick,anchor=west]
\tikzstyle{selected}=[draw=red,fill=red!30]
\tikzstyle{optional}=[dashed,fill=gray!50]
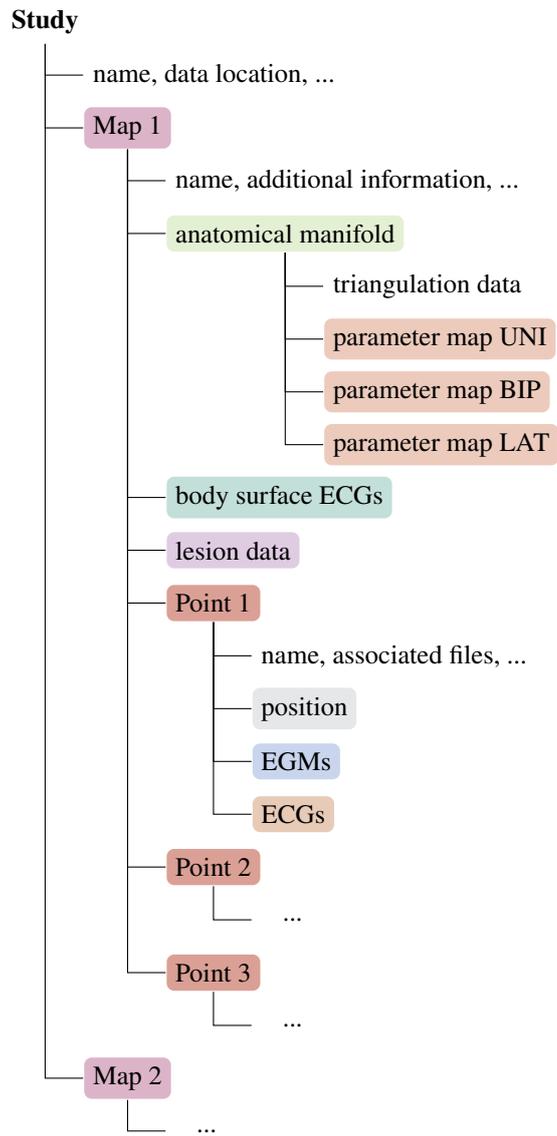
\begin{figure}[h!tb]
    \begin{tikzpicture}[%
      grow via three points={one child at (0.5,-0.7) and
      two children at (0.5,-0.7) and (0.5,-1.4)},
      edge from parent path={(\tikzparentnode.south) |- (\tikzchildnode.west)},
      study/.style={rectangle, draw=none, rounded corners=1mm, fill=white, text centered, anchor=west, text=black, minimum height=0.5em, minimum width =3em, font=\bfseries},
      map/.style={rectangle, draw=none, rounded corners=1mm, fill=RedViolet!30, text centered, anchor=west, text=black, minimum height=0.5em, minimum width =3em},
      point/.style={rectangle, draw=none, rounded corners=1mm, fill=Maroon!40, text centered, anchor=west, text=black, minimum height=0.5em, minimum width =3em},
      bsecg/.style={rectangle, draw=none, rounded corners=1mm, fill=PineGreen!20, text centered, anchor=west, text=black, minimum height=0.5em, minimum width =3em},
      ecg/.style={rectangle, draw=none, rounded corners=1mm, fill=RawSienna!20, text centered, anchor=west, text=black, minimum height=0.5em, minimum width =3em},
      egm/.style={rectangle, draw=none, rounded corners=1mm, fill=NavyBlue!20, text centered, anchor=west, text=black, minimum height=0.5em, minimum width =3em},
      surf/.style={rectangle, draw=none, rounded corners=1mm, fill=LimeGreen!20, text centered, anchor=west, text=black, minimum height=0.5em, minimum width =3em},
      param/.style={rectangle, draw=none, rounded corners=1mm, fill=Mahogany!20, text centered, anchor=west, text=black, minimum height=0.5em, minimum width =3em},
      pts/.style={rectangle, draw=none, rounded corners=1mm, fill=Gray!20, text centered, anchor=west, text=black, minimum height=0.5em, minimum width =3em},
      lesion/.style={rectangle, draw=none, rounded corners=1mm, fill=Plum!20, text centered, anchor=west, text=black, minimum height=0.5em, minimum width =3em},
      item/.style={rectangle, draw=none, rounded corners=1mm, fill=white, text centered, anchor=west, text=black, minimum height=0.5em, minimum width =3em},
      ],
      \node [study] {Study}
        child { node [item] {name, data location, ...}}		
        child { node [map] {Map 1}
          child { node [item] {name, additional information, ...}}
          child { node [surf] {anatomical manifold}
            child { node [item] {triangulation data}}
            child { node [param] {parameter map UNI}}
            child { node [param] {parameter map BIP}}
            child { node [param] {parameter map LAT}}
          }
          child [missing] {}
          child [missing] {}
          child [missing] {}
          child [missing] {}
          child { node [bsecg] {body surface ECGs}}
          child { node [lesion] {lesion data}}
          child { node [point] {Point 1}
            child { node [item] {name, associated files, ...}}
            child { node [pts] {position}}
            child { node [egm] {EGMs}}
            child { node [ecg] {ECGs}}
          }
          child [missing] {}
          child [missing] {}
          child [missing] {}
          child [missing] {}
          child { node [point] {Point 2}
            child {node [item] {...}}
          }
          child [missing] {}
          child { node [point] {Point 3}
            child {node [item] {...}}
          }
        }
        child [missing] {}
        child [missing] {}
        child [missing] {}
        child [missing] {}
        child [missing] {}
        child [missing] {}
        child [missing] {}
        child [missing] {}
        child [missing] {}
        child [missing] {}
        child [missing] {}
        child [missing] {}
        child [missing] {}
        child [missing] {}
        child [missing] {}
        child [missing] {}
        child [missing] {}
        child { node [map] {Map 2}
          child { node [item] {...}}
        };
    \end{tikzpicture}
    \caption{Hierarchical data structure used in \pE. Colors correspond to exported files shown in Fig.\ref{fig:output_folder}.}
    \label{fig:data_structure}
\end{figure}

\clearpage
\pagebreak
\section{Exported data formats}
\label{app:exportformats}

\begin{figure}[h!tb]
    \tcbset{on line, boxsep=2pt, left=0pt, right=0pt, top=0pt, bottom=0pt, colframe=white, colback=white}
    \begin{multicols}{2}
    \noindent\tcbox[colback=RedViolet!30]{\cartoMap}\\
    \indent \tcbox[]{\cartoMap}.\tcbox[colback=Gray!20]{pc}.\tcbox[colback=Emerald!30]{pts} \\
    \indent \tcbox[]{\cartoMap}.\tcbox[colback=Gray!20]{ppc}.\tcbox[colback=Emerald!30]{pts} \\
    \indent \tcbox[]{\cartoMap}.\tcbox[colback=LimeGreen!20]{surf}.\tcbox[colback=Emerald!30]{pts} \\
    \indent \tcbox[]{\cartoMap}.\tcbox[colback=LimeGreen!20]{surf}.\tcbox[colback=Emerald!30]{elem} \\
    \indent \tcbox[]{\cartoMap}.\tcbox[colback=LimeGreen!20]{surf}.\tcbox[colback=Emerald!30]{vtk} \\
    \indent \tcbox[]{\cartoMap}.\tcbox[colback=PineGreen!20]{bsecg}.\tcbox[]{ccf}.\tcbox[colback=Fuchsia!20]{json} \\
    \indent \tcbox[]{\cartoMap}.\tcbox[colback=PineGreen!20]{bsecg}.\tcbox[]{median}.\tcbox[colback=Fuchsia!20]{json} \\
    \indent \tcbox[]{\cartoMap}.\tcbox[colback=PineGreen!20]{bsecg}.\tcbox[]{mse}.\tcbox[colback=Fuchsia!20]{json} \\
    \indent \tcbox[]{\cartoMap}.\tcbox[colback=Maroon!40]{ptdata}.\tcbox[]{UNI}.\tcbox[colback=Gray!20]{pc}.\tcbox[colback=Sepia!40]{dat} \\
    \indent \tcbox[]{\cartoMap}.\tcbox[colback=Maroon!40]{ptdata}.\tcbox[]{BIP}.\tcbox[colback=Gray!20]{pc}.\tcbox[colback=Sepia!40]{dat} \\
    \indent \tcbox[]{\cartoMap}.\tcbox[colback=Maroon!40]{ptdata}.\tcbox[]{LAT}.\tcbox[colback=Gray!20]{pc}.\tcbox[colback=Sepia!40]{dat} \\
    \indent \tcbox[]{\cartoMap}.\tcbox[colback=RawSienna!20]{ecg}.\tcbox[]{I}.\tcbox[colback=Gray!20]
    {pc}.\tcbox[colback=CadetBlue!50]{igb}\\
    \indent \tcbox[]{\cartoMap}.\tcbox[colback=RawSienna!20]{ecg}.\tcbox[]{II}.\tcbox[colback=Gray!20]{pc}.\tcbox[colback=CadetBlue!50]{igb}\\
    \indent \tcbox[]{\cartoMap}.\tcbox[colback=RawSienna!20]{ecg}.\tcbox[]{III}.\tcbox[colback=Gray!20]{pc}.\tcbox[colback=CadetBlue!50]{igb}\\
    \indent \tcbox[]{\cartoMap}.\tcbox[colback=RawSienna!20]{ecg}.\tcbox[]{aVL}.\tcbox[colback=Gray!20]
    {pc}.\tcbox[colback=CadetBlue!50]{igb}\\
    \indent \tcbox[]{\cartoMap}.\tcbox[colback=RawSienna!20]{ecg}.\tcbox[]{aVR}.\tcbox[colback=Gray!20]{pc}.\tcbox[colback=CadetBlue!50]{igb}\\
    \indent \\
    \noindent\tcbox[colback=RedViolet!30]{\emph{2-LV VT}}\\
    \indent ..
    \vfill\null
    \columnbreak
    \indent \\
    \indent \tcbox[]{\cartoMap}.\tcbox[colback=RawSienna!20]{ecg}.\tcbox[]{aVF}.\tcbox[colback=Gray!20]{pc}.\tcbox[colback=CadetBlue!50]{igb}\\
    \indent \tcbox[]{\cartoMap}.\tcbox[colback=RawSienna!20]{ecg}.\tcbox[]{V1}.\tcbox[colback=Gray!20]{pc}.\tcbox[colback=CadetBlue!50]{igb}\\
    \indent \tcbox[]{\cartoMap}.\tcbox[colback=RawSienna!20]{ecg}.\tcbox[]{V2}.\tcbox[colback=Gray!20]{pc}.\tcbox[colback=CadetBlue!50]{igb}\\
    \indent \tcbox[]{\cartoMap}.\tcbox[colback=RawSienna!20]{ecg}.\tcbox[]
    {V3}.\tcbox[colback=Gray!20]{pc}.\tcbox[colback=CadetBlue!50]{igb}\\
    \indent \tcbox[]{\cartoMap}.\tcbox[colback=RawSienna!20]{ecg}.\tcbox[]{V4}.\tcbox[colback=Gray!20]{pc}.\tcbox[colback=CadetBlue!50]{igb}\\
    \indent \tcbox[]{\cartoMap}.\tcbox[colback=RawSienna!20]{ecg}.\tcbox[]{V5}.\tcbox[colback=Gray!20]{pc}.\tcbox[colback=CadetBlue!50]{igb}\\
    \indent \tcbox[]{\cartoMap}.\tcbox[colback=RawSienna!20]{ecg}.\tcbox[]{V6}.\tcbox[colback=Gray!20]{pc}.\tcbox[colback=CadetBlue!50]{igb}\\
    \indent \tcbox[]{\cartoMap}.\tcbox[colback=NavyBlue!20]{egm}.\tcbox[]{UNI}.\tcbox[colback=Gray!20]{pc}.\tcbox[colback=CadetBlue!50]{igb}\\
    \indent \tcbox[]{\cartoMap}.\tcbox[colback=NavyBlue!20]{egm}.\tcbox[]{BIP}.\tcbox[colback=Gray!20]{pc}.\tcbox[colback=CadetBlue!50]{igb}\\
    \indent \tcbox[]{\cartoMap}.\tcbox[colback=NavyBlue!20]{egm}.\tcbox[]{REF}.\tcbox[colback=Gray!20]{pc}.\tcbox[colback=CadetBlue!50]{igb}\\
    \indent \tcbox[]{\cartoMap}.\tcbox[colback=Mahogany!20]{map}.\tcbox[]{UNI}.\tcbox[colback=Sepia!40]{dat}\\
    \indent \tcbox[]{\cartoMap}.\tcbox[colback=Mahogany!20]{map}.\tcbox[]{BIP}.\tcbox[colback=Sepia!40]{dat}\\
    \indent \tcbox[]{\cartoMap}.\tcbox[colback=Mahogany!20]{map}.\tcbox[]{LAT}.\tcbox[colback=Sepia!40]{dat}\\
    \indent \tcbox[]{\cartoMap}.\tcbox[colback=Plum!20]{lesions}.\tcbox[colback=Emerald!30]{pts} \\
    \indent \tcbox[]{\cartoMap}.\tcbox[colback=Plum!20]{lesions}.\tcbox[colback=Lavender!30]{VisitagRFI}.\tcbox[colback=Sepia!40]{dat} \\
    \end{multicols}
    \caption{Any data associated with a mapping procedure is archived in a separated folder. The color-coded filename parts indicate a consistent and concise naming scheme devised to emphasize files which may be visualized together.}
    \label{fig:output_folder}
\end{figure}

\end{document}